 \documentclass[aps,prc,twocolumn,balance,floatfix,showpacs,amsmath,amssymb,nofootinbib]{revtex4-2}

\usepackage[utf8]{inputenc}
\usepackage{braket, amsmath}
\usepackage{mathtools}
\usepackage{tikz}
\usetikzlibrary{quantikz}
\usepackage{adjustbox}
\usepackage{color}
\usepackage[]{xcolor}
\usepackage{graphicx}
\usepackage{siunitx}
\usepackage{appendix}
\usepackage{dcolumn}
\usepackage{bm}
\usepackage{array}
\usepackage[bookmarks=true]{hyperref}
\usepackage{multirow}

\usepackage{braket}
\usepackage{caption}
\usepackage{subcaption}
\usepackage{academicons}
\usepackage{multirow}
\usepackage{upgreek}
\newcommand{\orcid}[1]{\href{https://orcid.org/#1}{\includegraphics[width=10pt]{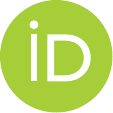}}}
\hypersetup{
	colorlinks=true,
	pdfborder={0 0 0},
	citecolor=blue,
	linkcolor=blue,
	urlcolor=blue
}

\begin{document}
	\title{Shape Transitions and Ground-State Properties of Tungsten Isotopes in Covariant Density Functional Theory}

	\author{Usuf Rahaman\orcid{0000-0003-0789-4257}}
	\email{urahaman@myamu.ac.in}
	\affiliation{%
		Department of Physics, Madanapalle Institute of Technology \& Science (MITS), Madanapalle-517325, India 
	}%

	\date{\today}

\begin{abstract}
	This study investigates the structural evolution of even-even tungsten isotopes ($^{154\text{--}264}$W) using covariant density functional theory (CDFT) with four relativistic functionals: DD-ME1, DD-ME2, DD-PC1, and DD-PCX. Key nuclear properties, including binding energies, quadrupole deformation parameters, two-neutron separation energies, neutron pairing energies, nuclear radii, and potential energy curves, are analyzed to explore shape transitions and stability from neutron-deficient to neutron-rich isotopes up to the drip line. The results reveal a dynamic shape evolution, with spherical configurations at $N = 82$ and $N = 126$, prolate dominance in intermediate regions, and shape coexistence in isotopes such as $^{158}$W, $^{160}$W, $^{194}$W, $^{196}$W, $^{206}$W, and near $^{244\text{--}248}$W. A potential subshell closure at $N = 118$ is identified, supported by anomalies in separation energies and vanishing pairing energies. The neutron drip line is predicted at $N = 184$, marked by a return to spherical symmetry. Comparisons with experimental data and other theoretical models, including the deformed Hartree-Fock-Bogoliubov method with the Skyrme SLy4 interaction, the Finite Range Droplet Model, and the Relativistic Mean Field model with NL3, show strong agreement, validating the robustness of CDFT. These findings enhance our understanding of nuclear structure in the medium-to-heavy mass region and provide insights relevant to r-process nucleosynthesis, thereby guiding future experimental studies at radioactive ion beam facilities.
	
	\vskip 0.2cm  
	
	
\end{abstract}

\maketitle

\section{Introduction}
\label{sec1}

The study of transitional nuclei constitutes a fundamental aspect of nuclear structure research, as it elucidates the quantum-mechanical mechanisms underlying the evolution of nuclear shapes from spherical to prolate and oblate configurations~\cite{Dennis2024, kiss2024, stev, Rahaman2025, Rahaman2024}.  
Transitional nuclei exhibit shape dynamics that reflect the interplay between collective and single-particle degrees of freedom, particularly in the medium-to-heavy mass region. In the atomic number region $Z = 72\text{--}80$, a rich variety of structural phenomena has been established, including shape coexistence, triaxiality, and quantum phase transitions~\cite{alkh, whel, john, pt1, pt2, cejnar2010}. 

Among these, tungsten ($Z = 74$) isotopes provide a particularly rich landscape for structural studies. 
Their properties are governed by a subtle balance between spherical shell effects and deformation-driving correlations, especially as the neutron number increases toward the drip line. Experimental and theoretical investigations have revealed ground-state shape transitions and possible shape coexistence in this mass region~\cite{rob2009, Sarriguren2008, Nomura2, Nomura3, Ramos2014}. 
Notably, isotopes around $A \sim 190$ lie near a critical point marking a transition between prolate and oblate shapes~\cite{sharma2025, dadwal2017, khalaf2016, jain2023, sharma2013, djerroud2000, barea2009, Rahaman2020, Usmani2018}. 
One of the key features in this mass region is the presence of shape coexistence, wherein multiple distinct shapes---such as spherical, prolate, or oblate---can coexist at nearly the same energy~\cite{morinaga1956, bonatsos2023}. 
These configurations are often separated by shallow potential energy surfaces, making shape mixing and quantum tunneling between them a topic of ongoing interest~\cite{john, Robert1991}. Tungsten isotopes are excellent candidates for exploring such phenomena due to their sensitivity to shell effects and collective correlations.

The structural evolution of tungsten isotopes is driven by variations in the neutron-to-proton ratio, which significantly influences nuclear deformation and collective behavior~\cite{karakatsanis2020}. Experimental studies, particularly those on neutron-rich tungsten isotopes, have identified a clear transition from spherical to deformed shapes along the isotopic chain. This is evidenced by changes in yrast-state half-lives, deformation parameters, and spectroscopic data~\cite{stevenson2005, mason2013, lane2010}.

Beyond nuclear structure, tungsten isotopes are significant in geochemistry, where their isotopic ratios in kimberlites and volcanic lavas provide constraints on mantle processes, plume evolution, and tectonics~\cite{nakanishi2021, mei2023, rubie2025, budde2016}. Thus, the relevance of tungsten isotopes spans both fundamental physics and Earth sciences.

Covariant Density Functional Theory (CDFT) stands out as a highly effective framework within nuclear density functional theory (DFT), leveraging Lorentz-covariant energy density functionals to explore nuclear properties across the nuclide chart~\cite{niksic2014,lalazissis2005,roca2011,niksic2002a,agbemava2014}. Its key strengths include a natural incorporation of spin-orbit coupling---critical for understanding nuclear shell structures---without the need for empirical tuning~\cite{ring1996,meng2006}. Additionally, CDFT provides a robust relativistic explanation of pseudospin symmetry through the interplay of scalar and vector potentials, accounting for nearly degenerate orbitals~\cite{liang2015}. By connecting to relativistic Brueckner-Hartree-Fock theory, CDFT is grounded in realistic nucleon-nucleon interactions~\cite{shen2019}. The framework also handles time-odd mean fields effectively, which are essential for modeling rotating nuclei, odd-mass systems, and magnetic properties~\cite{zhang2014, zhao2018}. Insights from nonrelativistic reductions of covariant functionals further strengthen its link to nonrelativistic models~\cite{ren2020}. As detailed in Ref.~\cite{RelDFTBook}, these features make CDFT exceptionally well-suited for studying ground-state properties, shape evolution, and a broad range of nuclear structure phenomena, including collective correlations and quantum phase transitions~\cite{abusara2012,agbemava2015,agbemava2017,meng2015,matev2007,afanasjev2008,afana2016}.

In this study, we employ the CDFT framework to systematically explore the ground-state properties of even-even tungsten isotopes ranging from neutron number $N = 80$ up to the predicted neutron drip line. Our analysis incorporates several state-of-the-art relativistic energy density functionals, including DD-ME1~\cite{niksic2002b}, DD-ME2~\cite{lalazissis2005}, DD-PC1~\cite{niksic2008}, and DD-PCX~\cite{yuksel2019}. We compute key observables such as binding energies, quadrupole deformation parameters, and charge radii. Particular attention is paid to identifying regions of shape transition, quantifying isotopic shifts, and assessing the robustness of different functionals. For benchmarking purposes, we compare our findings with results from the deformed Hartree-Fock-Bogoliubov (HFB) theory using the Skyrme SLy4 interaction in a transformed harmonic oscillator (THO) basis, the Finite Range Droplet Model (FRDM), the deformed relativistic Hartree-Bogoliubov theory in continuum (DRHBc) with the PC-PK1 functional, and the Relativistic Mean Field (RMF) approach with the NL3 parameter set. 

This investigation enhances our understanding of shape evolution and nuclear deformation in tungsten isotopes, providing essential insights for both nuclear theory and astrophysical modeling. The results are also relevant for interpreting r-process nucleosynthesis and for guiding future experimental investigations at radioactive ion beam facilities~\cite{doba1984, chabanat1998, erler2012}.

The paper is organized as follows: Section~\ref{sec2} describes the theoretical framework, Section~\ref{sec3} presents the results and discussion, and Section~\ref{sec4} provides the conclusions and future perspectives.

\section{Theoretical Framework}
\label{sec2}

\subsection{Covariant Density Functional Theory}

This study employs two formulations of Covariant Density Functional Theory (CDFT)~\cite{niksic2014, niksic2002a, cescato1998}: the density-dependent meson-exchange (DD-ME) model~\cite{lalazissis2005, niksic2002b} and the density-dependent point-coupling (DD-PC) model~\cite{niksic2008, yuksel2019}. These approaches differ fundamentally in the treatment of effective nucleon-nucleon interactions. The DD-ME model is based on finite-range meson exchange, whereas the DD-PC model replaces meson fields with zero-range contact interactions and derivative terms. Together, they offer complementary perspectives for assessing the sensitivity of nuclear observables to interaction range and parametrization choices.

\subsubsection{Meson-Exchange Model}

The meson-exchange model describes the nucleus as a system of Dirac nucleons interacting via finite-mass meson exchange~\cite{niksic2014,lalazissis2005, ring1996, niksic2002b, vretenar2005, typel1999, brockmann1992, hofmann2001}. The principal mesons mediating the interaction are the isoscalar-scalar $\sigma$ meson, the isoscalar-vector $\omega$ meson, and the isovector-vector $\rho$ meson. The corresponding Lagrangian density reads~\cite{gambhir1990, serot1986, serot1997}:

\begin{eqnarray}
	{\cal L} &=& \bar{\psi} \left[ \gamma^\mu (i\partial_\mu - g_{\omega}\omega_\mu - g_{\rho} \vec{\rho}_\mu \cdot \vec{\tau} - e A_\mu) - m - g_{\sigma}\sigma \right] \psi \nonumber \\
	&+& \frac{1}{2} (\partial_\mu \sigma)^2 - \frac{1}{2} m_{\sigma}^2 \sigma^2 - \frac{1}{4} \Omega_{\mu\nu} \Omega^{\mu\nu} + \frac{1}{2} m_{\omega}^2 \omega^\mu \omega_\mu \nonumber \\
	&-& \frac{1}{4} \vec{R}_{\mu\nu} \cdot \vec{R}^{\mu\nu} + \frac{1}{2} m_{\rho}^2 \vec{\rho}^\mu \cdot \vec{\rho}_\mu - \frac{1}{4} F_{\mu\nu} F^{\mu\nu}.
	\label{lagrangian_me}
\end{eqnarray}

Here, $\psi$ represents the nucleon Dirac spinor and $m$ is the bare nucleon mass. The field strength tensors for the vector fields are defined as:

\begin{equation}
	\Omega^{\mu\nu} = \partial^\mu \omega^\nu - \partial^\nu \omega^\mu, \\
	\vec{R}^{\mu\nu} = \partial^\mu \vec{\rho}^\nu - \partial^\nu \vec{\rho}^\mu, \\
	F^{\mu\nu} = \partial^\mu A^\nu - \partial^\nu A^\mu.
\end{equation}

The original Walecka model~\cite{walecka1974}, while foundational, was insufficient for precision nuclear structure studies owing to its overly stiff equation of state~\cite{boguta1977}. Modern meson-exchange models introduce either nonlinear meson self-interactions or density-dependent meson-nucleon couplings. In the former approach, a scalar self-interaction potential of the form

\begin{equation}
	U(\sigma) = \frac{1}{2} m_{\sigma}^2 \sigma^2 + \frac{1}{3} g_2 \sigma^3 + \frac{1}{4} g_3 \sigma^4
	\label{Usigma}
\end{equation}

replaces the bare mass term \(\frac{1}{2} m_{\sigma}^2 \sigma^2\) in the Lagrangian~\cite{reinhard1986, lalazissis1997, toddrutel2005}. In contrast, the density-dependent formalism used in this study introduces couplings of the form:

\begin{equation}
	g_i(\rho) = g_i(\rho_{\rm sat}) f_i(x), \quad x = \frac{\rho}{\rho_{\rm sat}}, \quad i = \sigma, \omega,
\end{equation}

where the above functional form is applied only to the isoscalar $\sigma$ and $\omega$ mesons. The function $f_i(x)$ is parameterized as

\begin{equation}
	f_i(x) = a_i \frac{1 + b_i (x + d_i)^2}{1 + c_i (x + d_i)^2}, \quad i = \sigma, \omega.
	\label{fx_me}
\end{equation}

For the isovector $\rho$ meson, a different density dependence is employed, given by the exponential form

\begin{equation}
	g_{\rho}(\rho) = g_{\rho}(\rho_{\rm sat}) \exp(-a_{\rho} (x - 1)).
\end{equation}

In this study, we use the DD-ME1~\cite{niksic2002b} and DD-ME2~\cite{lalazissis2005} parameter sets, with the corresponding parameters listed in Table~\ref{table:1}.

\begin{table}[ht]
	\caption{Parameters of the DD-ME1~~\cite{niksic2002b} and DD-ME2~~\cite{lalazissis2005} parameterizations for the meson-exchange model.}
	\centering
	\begin{tabular}{c c c}
		\hline
		Parameter & DD-ME1~~\cite{niksic2002b} & DD-ME2~~\cite{lalazissis2005} \\
		\hline
		$m$            & 939        & 939       \\
		$m_{\sigma}$   & 549.5255   & 550.1238  \\
		$m_{\omega}$   & 783        & 783       \\
		$m_{\rho}$     & 763        & 763       \\
		$g_{\sigma}$   & 10.4434    & 10.5396   \\
		$g_{\omega}$   & 12.8939    & 13.0189   \\
		$g_{\rho}$     & 3.8053     & 3.6836    \\
		$a_{\sigma}$   & 1.3854     & 1.3881    \\
		$b_{\sigma}$   & 0.9781     & 1.0943    \\
		$c_{\sigma}$   & 1.5342     & 1.7057    \\
		$d_{\sigma}$   & 0.4661     & 0.4421    \\
		$a_{\omega}$   & 1.3879     & 1.3892    \\
		$b_{\omega}$   & 0.8525     & 0.924     \\
		$c_{\omega}$   & 1.3566     & 1.462     \\
		$d_{\omega}$   & 0.4957     & 0.4775    \\
		$a_{\rho}$     & 0.5008     & 0.5647    \\
		\hline
	\end{tabular}
	\label{table:1}
\end{table}

\subsubsection{Point-Coupling Model}

The point-coupling formulation replaces meson exchange with contact interactions among nucleons~\cite{nikolaus1992, rusnak1997, burvenich2002}. The Lagrangian reads:

\begin{eqnarray}
	{\cal L} &=& \bar{\psi} (i \gamma^\mu \partial_\mu - m) \psi 
	- \frac{1}{2} \alpha_S(\hat{\rho}) (\bar{\psi} \psi)^2 
	- \frac{1}{2} \alpha_V(\hat{\rho}) (\bar{\psi} \gamma^\mu \psi)^2 \nonumber \\
	&-& \frac{1}{2} \alpha_{TV}(\hat{\rho}) (\bar{\psi} \vec{\tau} \gamma^\mu \psi)^2 
	- \frac{1}{2} \delta_S (\partial_\nu \bar{\psi} \psi)(\partial^\nu \bar{\psi} \psi) \nonumber \\
	&-& e \bar{\psi} \gamma^\mu A_\mu \frac{(1 - \tau_3)}{2} \psi.
	\label{lag_pc}
\end{eqnarray}

We adopt the DD-PC1~\cite{niksic2008} and DD-PCX~\cite{yuksel2019} parameterizations, summarized in Table~\ref{table:2}.

\begin{table}[ht]
	\caption{Parameters of the DD-PC1~~\cite{niksic2008} and DD-PCX~~\cite{yuksel2019} parameterizations for the point-coupling model.}
	\centering
	\begin{tabular}{c c c}
		\hline
		Parameter & DD-PC1~~\cite{niksic2008} & DD-PCX~~\cite{yuksel2019} \\
		\hline
		$m$ (MeV)      & 939         & 939         \\
		$a_s$ (fm$^2$) & -10.04616   & -10.97924   \\
		$b_s$ (fm$^2$) & -9.15042    & -9.03825    \\
		$c_s$ (fm$^2$) & -6.42729    & -5.31301    \\
		$d_s$          & 1.37235     & 1.37909     \\
		$a_v$ (fm$^2$) & 5.91946     & 6.43014     \\
		$b_v$ (fm$^2$) & 8.8637      & 8.87063     \\
		$d_v$          & 0.65835     & 0.65531     \\
		$b_{tv}$ (fm$^2$) & 1.83595  & 2.96321     \\
		$d_{tv}$       & 0.64025     & 1.30980     \\
		$\delta_s$ (fm$^4$) & -0.8149 & -0.87885    \\
		\hline
	\end{tabular}
	\label{table:2}
\end{table}

\subsection{Pairing Formalism}
\label{subsec:pairing}

Pairing correlations significantly influence the structure of open-shell and neutron-rich nuclei, especially those near the drip line~\cite{tanihata1996, Horowitz2001}. Within the CDFT framework, the Relativistic Hartree-Bogoliubov (RHB) model is employed to treat these correlations, seamlessly integrating mean-field and pairing effects in a unified framework~\cite{ring1996, Kucharek1991}.

The Bogoliubov quasiparticle creation operator is defined as:

\begin{equation}
	\alpha_k^\dagger = \sum_n U_{nk} c_n^\dagger + V_{nk} c_n,
\end{equation}

where $U_{nk}$ and $V_{nk}$ denote the components of the quasiparticle wave functions. The total energy is given by

\begin{equation}
	E_{\text{RHB}}[\hat{\rho}, \hat{\kappa}] = E_{\text{RMF}}[\hat{\rho}] + E_{\text{pair}}[\hat{\kappa}],
\end{equation}

with $\hat{\rho}$ the normal density and $\hat{\kappa}$ the pairing tensor. The pairing contribution is  

\begin{equation}
	E_{\text{pair}}[\hat{\kappa}] = \frac{1}{4} \sum_{n_1 n_1'} \sum_{n_2 n_2'} \kappa^*_{n_1 n_1'} \langle n_1 n_1' | V^{pp} | n_2 n_2' \rangle \kappa_{n_2 n_2'}.
\end{equation}

We use a separable finite-range pairing force~\cite{Tian2009,TIAN2009}, defined in coordinate space as:

\begin{equation}
	V^{pp}(\mathbf{r}_1, \mathbf{r}_2, \mathbf{r}_1', \mathbf{r}_2') = 
	-G \delta(\mathbf{R} - \mathbf{R}') P(\mathbf{r}) P(\mathbf{r}'),
\end{equation}

where $\mathbf{R} = \frac{1}{\sqrt{2}}(\mathbf{r}_1 + \mathbf{r}_2)$ and $\mathbf{r} = \frac{1}{\sqrt{2}}(\mathbf{r}_1 - \mathbf{r}_2)$, with $P(\mathbf{r})$ taken as a Gaussian:

\begin{equation}
	P(\mathbf{r}) = \frac{1}{(4\pi a^2)^{3/2}} \exp\left(-\frac{\mathbf{r}^2}{2a^2}\right).
\end{equation}

The pairing strength is taken as $G = 728$ MeV$\cdot$fm$^3$ with a range parameter $a = 0.644$ fm, calibrated to reproduce the pairing gap of the Gogny D1S interaction in symmetric nuclear matter~\cite{Berger1991}.

\subsection{Computational Details and Convergence}
\label{subsec:computational}
\begin{figure*}[ht]
	\centering
	\includegraphics[scale=0.5]{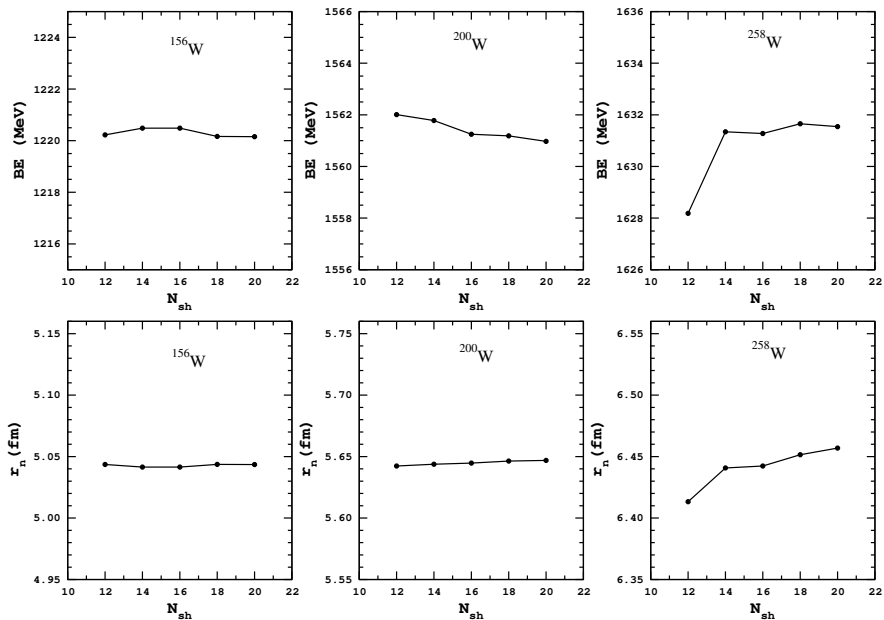}
	\caption{Binding energies (BE, top panels) and neutron root-mean-square (rms) radii (bottom panels) calculated within the CDFT+DD-ME2 framework for $^{156}\mathrm{W}$, $^{200}\mathrm{W}$, and $^{258}\mathrm{W}$ as functions of the number of oscillator shells $N_{\mathrm{sh}}$, illustrating the convergence of the basis.}
	\label{shell}
\end{figure*}

To handle weakly bound states and continuum effects in neutron-rich nuclei, we employ a harmonic oscillator (HO) basis expansion method. This approach is well established for relativistic Hartree-Bogoliubov calculations and provides an efficient and accurate description of both bound and low-lying continuum states through a discrete basis representation~\cite{eladri2020epjp}.

The RHB equations are solved by expanding the Dirac spinors and Bogoliubov quasiparticle wave functions in a set of HO basis states. The basis is characterized by the number of major oscillator shells \(N_{\text{sh}}\). For all calculations presented in this work, we use \(N_{\text{sh}} = 20\) major shells, which has been shown to be sufficient for heavy nuclei up to the neutron drip line.

To verify the adequacy of this basis size for the extremely neutron-rich tungsten isotopes considered in this study, we performed a systematic convergence test for the representative nuclei \(^{156}\mathrm{W}\), \(^{200}\mathrm{W}\), and \(^{258}\mathrm{W}\), covering the light, medium-mass, and neutron-rich regions of the isotopic chain. The variation of the binding energy and neutron rms radius with increasing basis size is presented in Fig.~\ref{shell}. 
Calculations were performed with \(N_{\text{sh}} = 12, 14, 16, 18,\) and \(20\) shells using the DD-ME2 functional.
Both observables exhibit clear saturation at the adopted shell numbers, confirming that the chosen configuration space is sufficient to ensure numerically reliable results across the entire tungsten isotopic chain.

\section{Results and Discussion}
\label{sec3}

\subsection{Binding Energy}

\begin{figure*}[ht]
	\centering
	\includegraphics[scale=0.5]{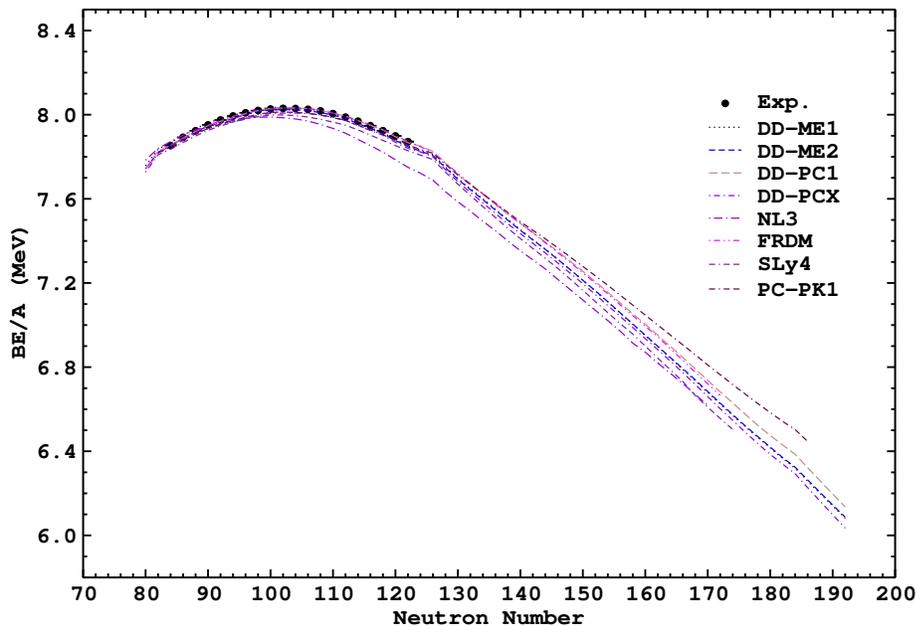}
	\caption{Binding energy per nucleon as a function of neutron number for even-even tungsten isotopes ($^{154\text{--}264}$W), calculated using CDFT with DD-ME1, DD-ME2, DD-PC1, and DD-PCX interactions. Theoretical results are compared with predictions from RMF (NL3), DRHBc (PC-PK1), HFB (SLy4), FRDM, and experimental data~\cite{wang2021}.}
	\label{Wbe}
\end{figure*}

The binding energy per nucleon is a fundamental quantity that provides insight into the stability and structural evolution of atomic nuclei. Figure~\ref{Wbe} displays the calculated ground-state binding energy per nucleon for even-even tungsten isotopes ($^{154\text{--}264}$W) using CDFT with four widely used interactions: DD-ME1~\cite{niksic2002b}, DD-ME2~\cite{lalazissis2005}, DD-PC1~\cite{niksic2008}, and DD-PCX~\cite{yuksel2019}.

Since experimental information is scarce for neutron-rich isotopes far from stability, comparison with other theoretical models is essential. We therefore compare our results with those obtained from the deformed HFB theory using the Skyrme SLy4 interaction with a THO basis~\cite{stoitsov2003}, the FRDM~\cite{moller2016}, the DRHBc with the PC-PK1 functional~\cite{Zhao2010,zhang2022,guo2024}, and the RMF model with the NL3 parameter set~\cite{mahapatro2015}. These cross-model benchmarks provide valuable insight into the reliability of CDFT across the isotopic chain.

The calculated binding energy trends show systematic behavior across the W isotopes, with a gradual increase in BE/A as the neutron number increases, reaching a peak around $A = 174$ ($N = 100$), after which the binding energy decreases. This observation is consistent across all four density functionals and aligns well with the trends observed in other theoretical predictions. The peak near $N = 100$ suggests enhanced stability, identifying $^{174}$W as the most bound isotope within the chain---a result consistent with both experimental systematics and other theoretical models.

Notably, the FRDM and SLy4-based HFB calculations show excellent agreement with our CDFT predictions, particularly on the neutron-rich side. The DRHBc (PC-PK1) results tend to yield slightly higher binding energies for the heavier isotopes, whereas RMF (NL3) calculations systematically underbind. Overall, the agreement among the different models and with available experimental values~\cite{wang2021} reinforces the robustness of the CDFT framework in describing heavy isotopes.

\subsection{Quadrupole Deformation}

\begin{figure*}[ht]
	\centering
	\includegraphics[scale=0.5]{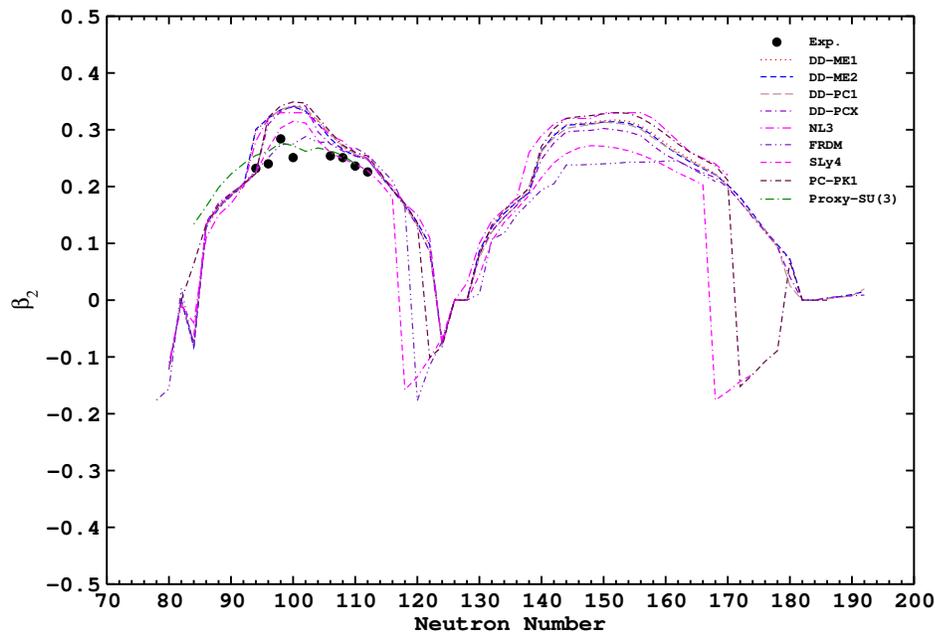}
	\caption{Quadrupole deformation parameter $\beta_2$ as a function of neutron number for even-even tungsten isotopes. Results from CDFT using DD-ME1, DD-ME2, DD-PC1, and DD-PCX interactions are compared with predictions from RMF (NL3), DRHBc (PC-PK1), HFB (SLy4), FRDM, the Proxy-SU(3) symmetry model, and available experimental data~\cite{raman2001, pritychenko2016}.}
	\label{Wbeta}
\end{figure*}

Nuclear deformation is an essential characteristic that plays a vital role in determining various nuclear observables, such as charge radii, transition probabilities, and electric quadrupole moments. The evolution of nuclear shapes across the tungsten isotopic chain is illustrated in Fig.~\ref{Wbeta}, which shows the calculated quadrupole deformation parameter ($\beta_2$) as a function of neutron number.

To gain a broader understanding of shape transitions, we compare our CDFT-based results using DD-ME1~\cite{niksic2002b}, DD-ME2~\cite{lalazissis2005}, DD-PC1~\cite{niksic2008}, and DD-PCX~\cite{yuksel2019} with predictions from several other theoretical models. These include the RMF model with the NL3 interaction~\cite{mahapatro2015}, DRHBc with the PC-PK1 functional~\cite{zhang2022,guo2024}, the HFB model with the Skyrme SLy4 force~\cite{stoitsov2003}, the Proxy-SU(3) symmetry model~\cite{bonatsos2023b, abuawwad2020, alstaty2022, abusara2023}, and the FRDM~\cite{moller2016}. Available experimental values of deformation~\cite{raman2001, pritychenko2016} are also included for validation.

Our analysis reveals a rich structural evolution across the W isotopic chain. Starting from neutron-deficient nuclei, the isotopes exhibit a predominantly prolate shape. As neutrons are added, a transition toward oblate and spherical configurations is observed around $N \approx 116$--$128$, followed by a return to prolate deformation in the heavier neutron-rich isotopes. This behavior underscores the transitional nature of tungsten isotopes and reflects subtle changes in shell structure and pairing correlations.

The calculated $\beta_2$ values from the four CDFT interactions display good consistency with available experimental data and with other theoretical predictions across most of the isotopic range. However, minor discrepancies appear in the neutron-rich region ($N = 116$--$122$), particularly in comparison with the SLy4-based HFB, DRHBc (PC-PK1), Proxy-SU(3) symmetry model, and FRDM predictions. While most models, including our CDFT calculations, indicate a weakly prolate shape near the drip line, the DRHBc (PC-PK1) and SLy4-based HFB models occasionally predict a slight oblate deformation in this region. These differences highlight the sensitivity of deformation predictions to the choice of functional and model space.

\subsection{Two-Neutron Separation Energy ($S_{2n}$), Shell Gap ($\delta S_{2n}$), and Neutron Pairing Energy ($E_{\text{pair},n}$)}

Understanding shell closures is fundamental for elucidating the underlying structure of nuclei. The two-neutron separation energy, defined as
\begin{equation}
	S_{2n}(N,Z) = B(N,Z) - B(N-2,Z),
\end{equation}
and the two-neutron shell gap, defined as
\begin{equation}
	\delta S_{2n}(N,Z) = S_{2n}(N,Z) - S_{2n}(N+2,Z),
\end{equation}
are key observables employed to probe such closures. 
A pronounced drop in $S_{2n}$ or a sharp peak in $\delta S_{2n}$ is generally regarded as a signature of a shell closure; however, local irregularities may also arise from shape coexistence or rapid changes in deformation and therefore must be corroborated by additional observables.

\begin{figure*}[ht]
	\centering
	\includegraphics[scale=0.5]{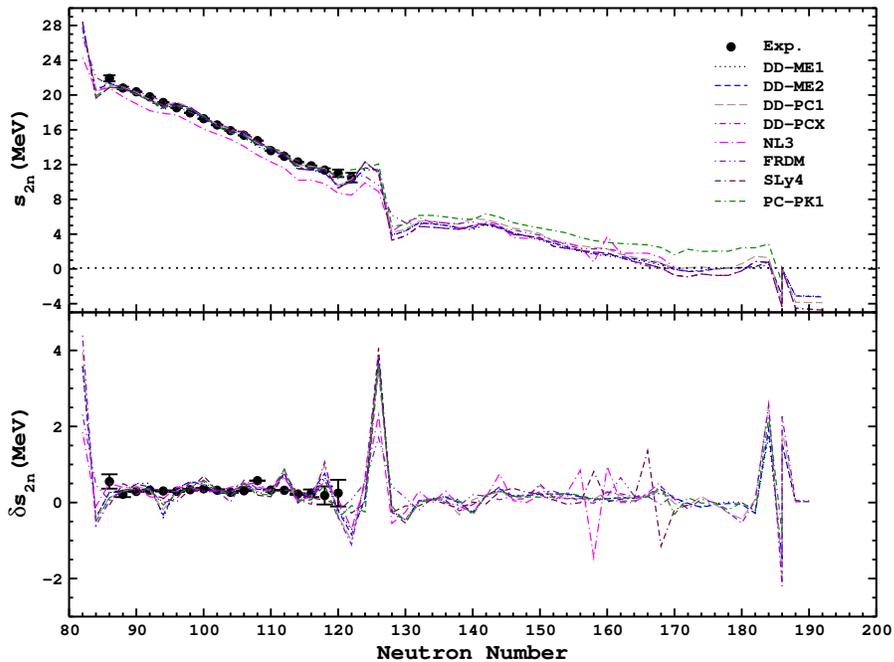}
	\caption{Two-neutron separation energy ($S_{2n}$, top panel) and two-neutron shell gap ($\delta S_{2n}$, bottom panel) as functions of neutron number for even-even W isotopes calculated using CDFT with DD-ME1, DD-ME2, DD-PC1, and DD-PCX interactions. Results are compared with experimental data, RMF (NL3), DRHBc (PC-PK1), HFB (SLy4), and FRDM predictions.}
	\label{Wsn}
\end{figure*}

Figure~\ref{Wsn} presents the calculated values of $S_{2n}$ and $\delta S_{2n}$ for even-even $^{154\text{--}264}$W isotopes using the CDFT framework with four density-dependent interactions: DD-ME1, DD-ME2, DD-PC1, and DD-PCX. These results are compared with experimental data and theoretical predictions from the RMF model with the NL3 interaction~\cite{mahapatro2015}, the DRHBc model with PC-PK1~\cite{zhang2022,guo2024}, the HFB approach using the Skyrme SLy4 force~\cite{stoitsov2003}, and the FRDM~\cite{moller2016}.

All four CDFT interactions predict the neutron drip line to be at $N = 184$; for $^{258}$W ($N=184$) the two-neutron separation energy is still positive but very small, and it becomes negative for the next isotope, indicating the limit of bound nuclei~\cite{eladri2020}. This estimate is consistent with the RMF (NL3), HFB (SLy4), DRHBc (PC-PK1), and FRDM predictions.

As evident from Fig.~\ref{Wsn}, $S_{2n}$ decreases smoothly with increasing neutron number, reflecting the gradual filling of neutron orbitals. Clear discontinuities in $S_{2n}$ and corresponding pronounced peaks in $\delta S_{2n}$ are observed at $N = 82$, $N = 126$, and $N = 184$, confirming these well-established magic numbers across all theoretical models~\cite{siddiqui2020, sharma2025b} and experimental data.
Beyond these classic shell closures, smaller kinks appear around $N = 112$ and $N = 118$ in the CDFT calculations. While such features may indicate subshell effects, they can also originate from competing prolate--oblate configurations and deformation-driven level rearrangements, and therefore should not be interpreted as definitive evidence of new magic numbers without additional support.

In this respect, the magnitude of the $\delta S_{2n}$ peak at $N = 118$ is systematically larger than that at $N = 112$ for most functionals, indicating a stronger gap at $N = 118$.

A comparison with the FRDM predictions shows smoother behavior in this region, suggesting that the appearance and strength of these subshell features are model dependent, as has been observed in other isotopic chains.

We note that $S_{2n}$, $\delta S_{2n}$, and pairing energies can be influenced by the adopted pairing strength and quasiparticle energy cutoff. In the present work, however, the same pairing functional, strength, and cutoff scheme are applied consistently for all isotopes and interactions, so that the relative trends with neutron number remain robust.

Although local structures are visible at both $N=112$ and $N=118$, their magnitudes differ. The peak in $\delta S_{2n}$ at $N=118$ is systematically larger for all four functionals, whereas the structure at $N=112$ is weaker and more model dependent. Therefore, $N=112$ is treated as a tentative subshell feature rather than a robust semimagic number.

To further investigate shell effects, we analyze the neutron pairing energy ($E_{\text{pair},n}$), which typically approaches zero at shell closures~\cite{delestal2001, sil2004}. 
The vanishing of pairing correlations at a given neutron number provides an independent and sensitive indicator of a shell gap, complementing the information obtained from $S_{2n}$ and $\delta S_{2n}$.
As illustrated in Fig.~\ref{Wep}, $E_{\text{pair},n}$ vanishes at $N = 82$, $N = 126$, and $N=184$, in alignment with the shell gaps observed in $S_{2n}$ and $\delta S_{2n}$. 
In the neutron-rich region, a strong suppression of pairing is also obtained at $N = 118$ for all four interactions, whereas at $N = 112$ the pairing energy exhibits only a shallow minimum and does not reach zero for some functionals, indicating a weaker subshell character.
At $N=112$, the neutron pairing energy exhibits only a shallow minimum and does not vanish for some interactions, in contrast to the strong suppression obtained at $N=118$. This behavior indicates that only $N=118$ satisfies the simultaneous criteria of a pronounced $\delta S_{2n}$ peak and near-vanishing pairing energy.

This behavior is analogous to situations reported in other regions of the nuclear chart, where the simultaneous occurrence of a $\delta S_{2n}$ peak and vanishing pairing energy provides robust evidence for a shell closure.

\begin{figure*}[ht]
	\centering
	\includegraphics[scale=0.5]{epair_W.eps}
	\caption{Neutron pairing energy ($E_{\text{pair},n}$) as a function of neutron number for even-even W isotopes calculated with CDFT using DD-ME1, DD-ME2, DD-PC1, and DD-PCX. Vanishing or near-zero values at $N = 82$, $N = 126$, and a pronounced minimum at $N = 118$ suggest shell and subshell closures. A complete vanishing is also observed at $N = 184$ (see Fig.~\ref{Wsn}), further confirming the magic character of this neutron number.}
	\label{Wep}
\end{figure*}

Taken together, the combined analysis of $S_{2n}$, $\delta S_{2n}$, and $E_{\text{pair},n}$ indicates that $N=118$ emerges as the most consistent candidate for a neutron subshell closure in neutron-rich tungsten isotopes, whereas the structure at $N=112$ appears to be weaker and more model dependent in character.

\subsection{Neutron, Proton, and Charge Radii}

\begin{figure*}[ht]
	\centering
	\includegraphics[scale=0.5]{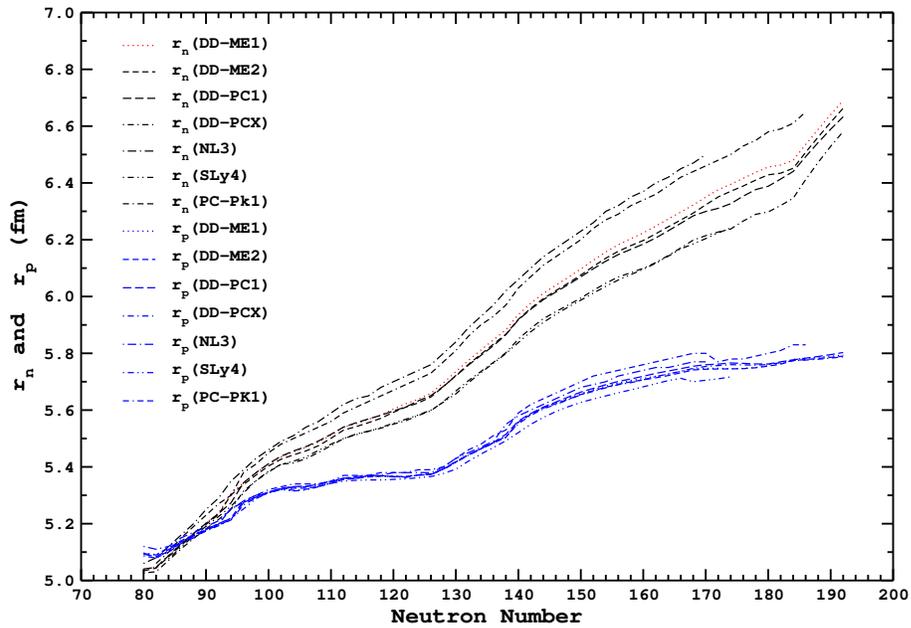}
	\caption{RMS neutron radius ($r_n$) and proton radius ($r_p$) for even-even W isotopes as calculated with CDFT using four density-dependent interactions. Results are compared with RMF (NL3), DRHBc (PC-PK1), and HFB (SLy4) models. Discontinuities at $N = 82$, $N = 126$, and $N = 184$ reflect closed-shell behavior.}
	\label{Wrad}
\end{figure*}

Nuclear radii provide essential information on the spatial distribution of nucleons. A particularly important observable is the neutron skin thickness, defined as the difference between the neutron and proton root-mean-square (RMS) radii. This quantity becomes more pronounced in neutron-rich systems and is strongly linked to the isospin dependence of the nuclear equation of state, thereby connecting the structure of finite nuclei with properties of neutron matter~\cite{reinhard2016,hagen2016,agrawal2006}.

Figure~\ref{Wrad} shows the evolution of neutron ($r_n$) and proton ($r_p$) RMS radii for even-even tungsten isotopes, calculated using the CDFT framework with DD-ME1, DD-ME2, DD-PC1, and DD-PCX interactions. As expected, $r_n$ increases steadily with neutron number, while distinct discontinuities at $N = 82$, $N = 126$, and $N = 184$ indicate shell closures. In contrast, $r_p$ increases only very gradually with neutron number; the slight enlargement of approximately $0.1\,\mathrm{fm}$ can be attributed to the increased occupation of neutron orbitals that polarize the proton core and to the general increase of the nuclear volume with mass number.

Differences among theoretical models emerge particularly in the neutron-rich region. For instance, RMF (NL3) and DRHBc with PC-PK1 tend to predict slightly larger $r_n$ and $r_p$ values than other models, indicating a stiffer isovector interaction. On the other hand, the HFB (SLy4) model predicts more compact neutron radii for heavier isotopes.

\begin{figure*}[ht]
	\centering
	\includegraphics[scale=0.5]{rc_W.eps}
	\caption{Root-mean-square (RMS) charge radii ($r_c$) of even-even W isotopes obtained from CDFT calculations using DD-ME1, DD-ME2, DD-PC1, and DD-PCX interactions. Theoretical results are compared with available experimental data~\cite{angeli2004, angeli2013}, RMF (NL3), and DRHBc (PC-PK1) models.}
	\label{Wrc}
\end{figure*}

Figure~\ref{Wrc} shows the RMS charge radii ($r_c$) for W isotopes. Our CDFT results with all four interactions exhibit good agreement with experimental data~\cite{angeli2004, angeli2013}. RMF (NL3) and DRHBc (PC-PK1) slightly overestimate the charge radii for neutron-rich isotopes beyond $N > 140$, possibly due to differences in the treatment of pairing and deformation.

\begin{figure*}[ht]
	\centering
	\includegraphics[scale=0.5]{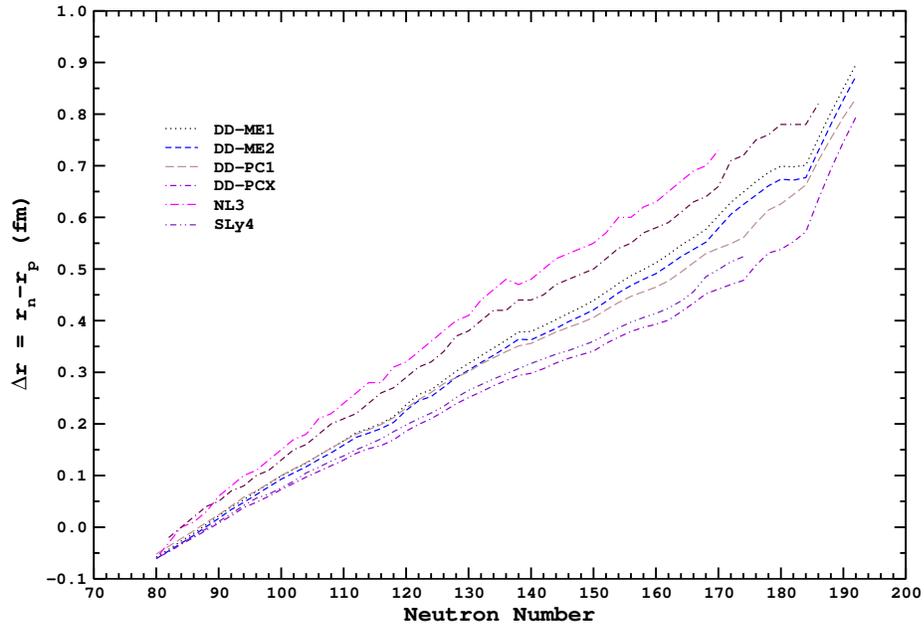}
	\caption{Neutron skin thickness ($\Delta r = r_n - r_p$) for even-even W isotopes as a function of neutron number. Results from CDFT with DD-ME1, DD-ME2, DD-PC1, and DD-PCX are compared with predictions from RMF (NL3), DRHBc (PC-PK1), and HFB (SLy4).}
	\label{Wskin}
\end{figure*}

Figure~\ref{Wskin} presents the neutron skin thickness $\Delta r$ as a function of neutron number. A nearly linear increase is observed across the isotopic chain, particularly beyond $N = 126$, reflecting the progressive development of a neutron-rich surface layer. The slope and magnitude of $\Delta r$ are sensitive to the isovector properties of the interactions used, with RMF (NL3) again showing the largest neutron skins.

\subsection{Potential Energy Curves}
\label{sec3.5}

The potential energy curves (PECs) are obtained by performing constrained CDFT calculations in which the axial mass quadrupole moment is fixed to selected values, allowing the total energy to be mapped as a function of the deformation parameter $\beta_2$. The constraint is implemented through the method of quadratic constraint~\cite{ring1980}, in which the quantity
\begin{equation}
	\langle \hat{H} \rangle +
	\sum_{\mu=0,2} C_{2\mu}\left(\langle \hat{Q}_{2\mu} \rangle - q_{2\mu}\right)^2
\end{equation}
is minimized self-consistently. Here, $\langle \hat{H} \rangle$ denotes the total energy of the system, $\langle \hat{Q}_{2\mu} \rangle$ are the expectation values of the mass quadrupole operators, $q_{2\mu}$ are the constrained multipole moments, and $C_{2\mu}$ are the corresponding stiffness constants.

The mass quadrupole operators are defined as
\begin{equation}
	\hat{Q}_{20} = 2z^2 - x^2 - y^2, 
	\qquad
	\hat{Q}_{22} = x^2 - y^2.
\end{equation}

The quadratic constraint introduces an additional term in the self-consistent equations,
\begin{equation}
	\sum_{\mu=0,2} \lambda_{\mu} \hat{Q}_{2\mu},
\end{equation}
where the Lagrange multipliers are given by
\begin{equation}
	\lambda_{\mu} = 2C_{2\mu}\left(\langle \hat{Q}_{2\mu} \rangle - q_{2\mu}\right).
\end{equation}
This term drives the system toward the desired point in deformation space and enables the exploration of configurations away from the stationary minimum.

By scanning a wide range of constrained quadrupole moments, the total energy surface is mapped and the locations of spherical, prolate, and oblate minima are identified. The global minimum determines the equilibrium shape, while additional local minima indicate possible shape coexistence. Convergence of the constrained solutions is ensured by employing the same optimized harmonic oscillator basis used for the ground-state calculations, with the tested number of oscillator shells for fermions and bosons. Calculations are performed using DD-ME1~\cite{niksic2002b}, DD-ME2~\cite{lalazissis2005}, DD-PC1~\cite{niksic2008}, and DD-PCX~\cite{yuksel2019}, and the resulting PECs for even-even tungsten isotopes are presented in Figs.~\ref{pw1}--\ref{pw3}.

\begin{figure*}[hbt!]
	\centering
	\includegraphics[scale=0.5]{pse1.eps}
	\caption{Potential energy curves for even-even $^{156\text{--}186}$W isotopes as functions of quadrupole deformation $\beta_2$, calculated using DD-ME1, DD-ME2, DD-PC1, and DD-PCX interactions. Energies are normalized to the ground state.}
	\label{pw1}
\end{figure*}

\begin{figure*}[hbt!]
	\centering
	\includegraphics[scale=0.5]{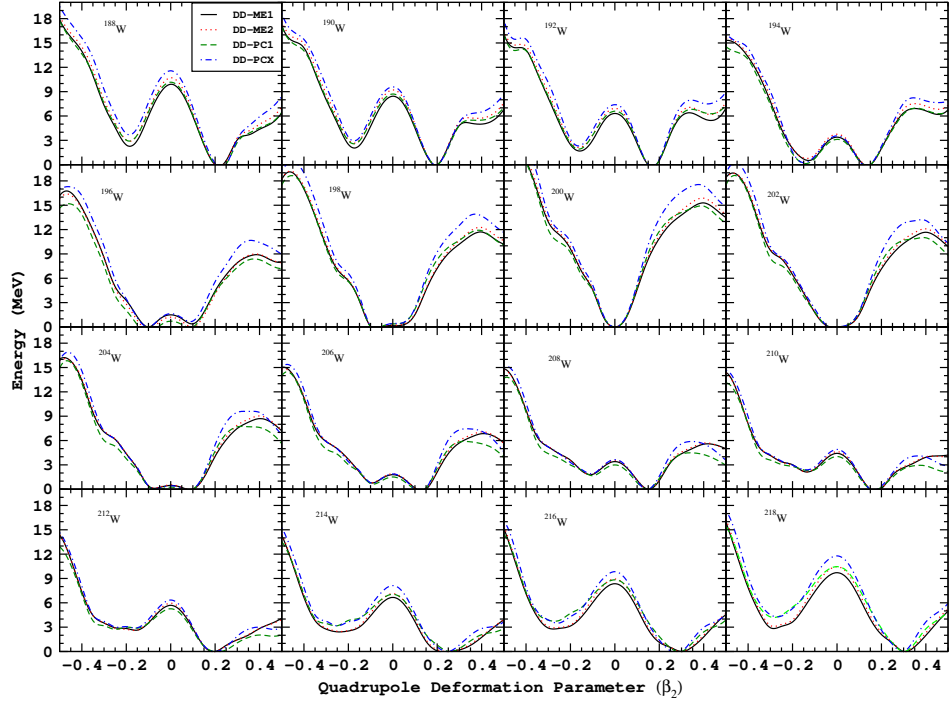}
	\caption{Same as Fig.~\ref{pw1}, but for $^{188\text{--}218}$W isotopes.}
	\label{pw2}
\end{figure*}

\begin{figure*}[hbt!]
	\centering
	\includegraphics[scale=0.5]{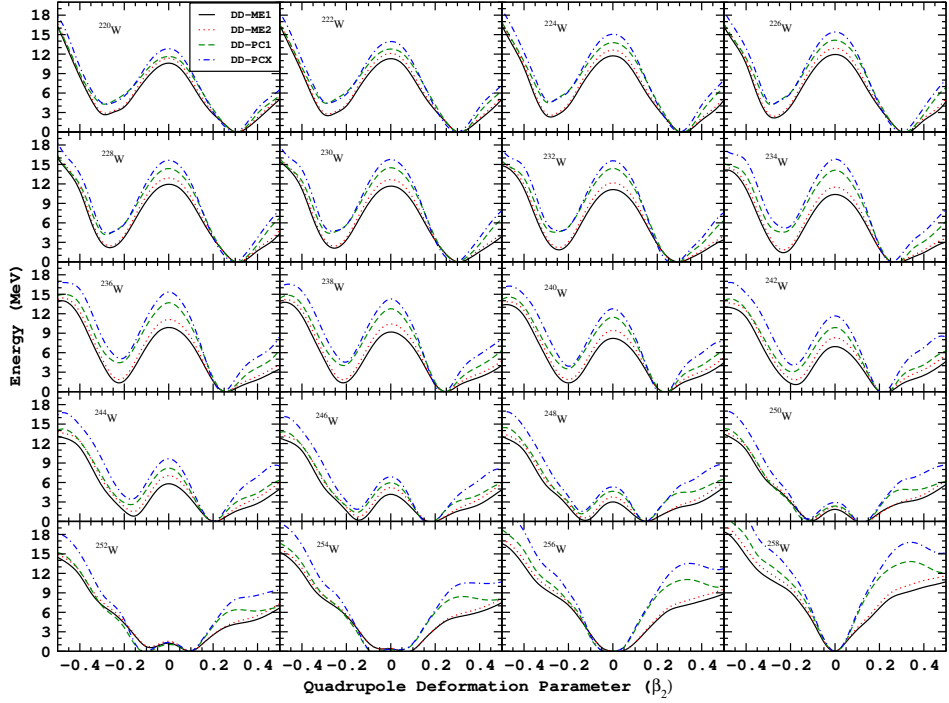}
	\caption{Same as Fig.~\ref{pw1}, but for $^{220\text{--}256}$W isotopes.}
	\label{pw3}
\end{figure*}

The PECs reveal a rich pattern of shape evolution across the W isotopic chain. At $N = 82$ ($^{156}$W), a pronounced spherical minimum is observed due to the shell closure. With the addition of neutrons, the equilibrium shape shifts rapidly toward prolate deformation. For isotopes such as $^{158}$W and $^{160}$W, both prolate and oblate shapes coexist, with their minima differing by less than 1 MeV, highlighting a strong competition between the two configurations.

From $^{162}$W to $^{192}$W, the prolate minimum remains the most energetically favored, while oblate configurations appear only as shallow secondary minima. A second region of shape coexistence emerges in $^{194}$W and $^{196}$W, where prolate and oblate minima again lie within 1 MeV of each other.

Near $N = 126$ ($^{200}$W), the ground state again becomes spherical, consistent with shell effects. The isotope $^{206}$W exhibits shape coexistence, with prolate and oblate minima separated by approximately 0.7 MeV. 
For even heavier isotopes, from $^{208}$W up to $^{242}$W, the PECs display well-deformed prolate minima that are typically $2.0$--$4.0$ MeV deeper than the oblate minima, depending on the interaction. This indicates robust prolate dominance in this region. Nevertheless, weak oblate configurations persist as higher-lying local minima, showing that the possibility of competing shapes is not entirely suppressed.

Interestingly, in the neutron-rich isotopes $^{244}$W, $^{246}$W, and $^{248}$W, shape coexistence reappears for some interactions (e.g., DD-ME1, DD-ME2), suggesting a soft potential energy surface and enhanced shape fluctuations characteristic of transitional nuclei.

In the very neutron drip-line region, particularly in $^{250}$W and $^{252}$W, shape coexistence reappears. Finally, at $N=184$ ($^{258}$W), the PECs reveal a sharp spherical minimum, confirming the re-emergence of sphericity due to the predicted neutron magic number.

Overall, the study reveals a rich pattern of shape evolution in tungsten isotopes: starting from spherical configurations at $N=82$, evolving through regions of shape coexistence and strong prolate dominance, regaining spherical symmetry at $N=126$, re-entering deformed regimes on the neutron-rich side, and finally returning to spherical symmetry at $N=184$.
Our results are consistent with previous studies on heavy deformed nuclei~\cite{Sarriguren2008, afana2016, choi2022} and reinforce the view that tungsten isotopes exhibit pronounced shape transitions, coexistence, and stabilization of prolate deformation in the neutron-rich domain.

The potential energy curves do not display a pronounced spherical minimum at either $N=112$ or $N=118$, suggesting that the observed gaps are associated with deformed subshell effects rather than strong spherical magicity. A definitive confirmation would require a detailed analysis of the single-particle level structure, ideally including tensor contributions~\cite{otsuka2020}.

The present calculations are restricted to axial symmetry in order to maintain a fully systematic and self-consistent description of the entire even-even tungsten isotopic chain within a uniform numerical framework.
It is well known that shallow competing minima in axial potential energy curves indicate possible $\gamma$ softness and the presence of triaxial correlations. In the present study, several nuclei exhibit prolate and oblate minima separated by less than approximately 1~MeV, suggesting a soft energy surface with respect to the triaxial degree of freedom. A full two-dimensional $(\beta,\gamma)$ constrained CDFT analysis would provide a more detailed characterization of these features.
However, such calculations require a substantially different computational framework and are beyond the scope of the present large-scale axial study.

\section{Conclusion}
\label{sec4}

This study systematically investigates the structural evolution of even-even tungsten isotopes ($^{154\text{--}264}$W) using the CDFT framework with four relativistic energy density functionals: DD-ME1, DD-ME2, DD-PC1, and DD-PCX. By analyzing key observables such as binding energies, quadrupole deformation parameters, two-neutron separation energies, neutron pairing energies, nuclear radii, and potential energy curves, we provide a comprehensive picture of shape transitions and nuclear stability across the isotopic chain, from neutron-deficient to neutron-rich regions up to the predicted drip line.

Our results reveal a complex pattern of shape evolution in tungsten isotopes. Spherical configurations dominate at neutron magic numbers $N = 82$, $N = 126$, and $N = 184$, while prolate deformation prevails in intermediate regions, with notable shape coexistence observed in isotopes such as $^{158}$W, $^{160}$W, $^{194}$W, $^{196}$W, $^{206}$W, and neutron-rich isotopes near $^{244\text{--}248}$W. The axial potential energy curves exhibit shallow competing minima in several transitional nuclei, indicating $\gamma$ softness and suggesting the possible presence of triaxial correlations. The neutron drip line is consistently predicted at $N = 184$, as the two-neutron separation energy becomes negative for the next isotope. These findings align well with available experimental data~\cite{wang2021, raman2001, pritychenko2016, angeli2004, angeli2013} and show strong agreement with predictions from other theoretical models, including the deformed Hartree-Fock-Bogoliubov (HFB) approach with Skyrme SLy4, the Finite Range Droplet Model (FRDM), the deformed relativistic Hartree-Bogoliubov theory in continuum (DRHBc) with PC-PK1, the Proxy-SU(3) symmetry model, and the Relativistic Mean Field (RMF) with NL3~\cite{stoitsov2003, moller2016, Zhao2010, zhang2022, guo2024, bonatsos2023b, mahapatro2015}. Minor discrepancies, particularly in deformation predictions near $N = 116\text{--}122$ and in neutron radii for heavier isotopes, highlight the sensitivity of these observables to the choice of functional and model space.

The analysis of two-neutron separation energies ($S_{2n}$) and shell gaps ($\delta S_{2n}$) confirms shell closures at $N = 82$, $N = 126$, and $N = 184$, while both $N = 112$ and $N = 118$ exhibit signatures of subshell structure through peaks in $\delta S_{2n}$ and vanishing neutron pairing energies. The larger and more persistent gap at $N = 118$ indicates that it represents the more robust subshell closure in neutron-rich tungsten isotopes. The neutron skin thickness increases nearly linearly with neutron number, particularly beyond $N = 126$, reflecting the development of a neutron-rich surface and providing insight into the isovector properties of the nuclear equation of state.

These findings enhance our understanding of nuclear structure in the medium-to-heavy mass region and have implications for astrophysical processes such as r-process nucleosynthesis, in which tungsten isotopes play a significant role. The robustness of CDFT across different functionals, combined with its consistency with experimental and theoretical benchmarks, establishes it as a reliable tool for studying nuclear properties near and beyond the limits of stability.

Future experimental validation of the predicted subshell features at $N = 112$ and $N = 118$, and of shape coexistence in neutron-rich isotopes, can be pursued at advanced radioactive ion beam facilities. Additionally, extending CDFT calculations to odd-mass tungsten isotopes and incorporating triaxial degrees of freedom could provide deeper insights into the complex dynamics of shape transitions. These efforts will further bridge nuclear physics with astrophysical modeling, advancing our knowledge of nuclear interactions and their role in the cosmos.
	
\section*{Acknowledgements}
Usuf Rahaman would like to acknowledge the Department of Physics, Aligarh Muslim University for computational support.



\end{document}